\documentclass{aa}  

\usepackage{graphicx}
\usepackage{txfonts}
\begin{document}

   \title{Re-analysis of the 267-GHz ALMA observations of Venus}

   \subtitle{No statistically significant detection of phosphine\thanks{Since the publication of GRB20, the authors alerted us about an update in the ALMA data processing script and made the new script available. In parallel, the data available in the ALMA Science Archive is undergoing (so-called QA3) reprocessing to include the same correction. The resulting reprocessed data no longer contains the strong ripples that GRB20 report and that we also find, for example as manifested in the non-Gaussian noise distribution. In these reprocessed data we do not find a clear absorption feature that can be attributed to PH$_3$, although further exploration of these data is necessary to analyse this in more detail.}}

   \author{I.A.G. Snellen
          \inst{1}
          L. Guzman-Ramirez \inst{1}
          M.R. Hogerheijde
          \inst{1,2}
          A.P.S. Hygate \inst{1}
          F.F.S. van der Tak \inst{3,4}
          }

   \institute{$^1$Leiden Observatory, Leiden University, Postbus 9513, 2300 RA Leiden, The Netherlands\\
   $^2$ Anton Pannekoek Institute for Astronomy, University of Amsterdam, Science Park 904, 1090 GE Amsterdam, The Netherlands\\
    $^3$ SRON Netherlands Institute for Space Research, Landleven 12, 9747 AD Groningen, The Netherlands\\
    $^4$ Kapteyn Astronomical Institute, University of Groningen, Landleven 12, 9747 AD Groningen, The Netherlands\\    }

   \date{}

  \abstract
   {ALMA observations of Venus at 267 GHz have been presented in the literature that show the apparent presence of phosphine (PH$_3$) in its atmosphere. Phosphine has currently no evident production routes on the planet's surface or in its atmosphere.}
   {The aim of this work is to assess the statistical reliability of the line detection by independent re-analysis of the ALMA data.}
   {The ALMA data were reduced as in the published study, following the provided scripts. First the spectral analysis presented in the study was reproduced and assessed.  Subsequently, the spectrum was statistically evaluated, including its dependence on selected ALMA baselines.}
   {We find that the 12$^{th}$-order polynomial fit to the spectral passband utilised in the published study leads to spurious results. Following their recipe, five other $>$10$\sigma$ lines can be produced in absorption or emission within 60 km s$^{-1}$ from the PH$_3$ 1-0 transition frequency by suppressing the surrounding noise. Our independent analysis shows a feature near the PH$_3$ frequency at a $\sim$2$\sigma$ level, below the common threshold for statistical significance. Since the spectral data have a non-Gaussian distribution, we consider a feature at such level as statistically unreliable that cannot be linked to a false positive probability.}
   {We find that the published 267-GHz ALMA data provide no statistical evidence for phosphine in the atmosphere of Venus.}

               
\titlerunning{No statistically significant detection of phosphine}
\authorrunning{Snellen et al.}
   \maketitle
%

\section{Introduction}
Recently, ALMA (Atacama Large Millimeter Array) observations of Venus at 267 GHz have been presented that show the apparent presence of phosphine (PH$_3$) in its atmosphere at $\sim$20 parts-per-billion (Greaves et al. 2020; GRB20 hereafter). Since phosphorus is expected to be in oxidised forms, phosphine has no easily explained production routes on the planet's surface or in its atmosphere at this level (Bains et al. 2020). At the same time, PH$_3$ is identified as a potential biomarker gas (Sousa-Silva et al. 2020), and an aerial biosphere of Venusian microbes (Seager et al. 2020) is being proposed as the possible source. The required biomass is potentially just a fraction of that of the Earth's aerial biosphere (Lingam \& Loeb 2020). Also, the Venusian life may have an Earth origin (Siraj \& Loeb 2020). A balloon mission is  proposed to search in situ for these lifeforms (Hein et al. 2020), which could already be launched in 2022-2023.  In the meantime, Encrenaz et al. (2020) have provided a stringent upper limit on the PH$_3$ abundance of $<$5 ppb from observations in the thermal infrared, which, in the absence of variability, is in conflict with the results presented by GRB20. 

The aim of this work is to assess the statistical reliability of the PH$_3$ J = $1-0$ line detection by independently re-analysing the ALMA data. In Section 2, the processing and calibration of the ALMA data is described, performed in a similar way as by GRB20. In Section 3, the procedure that led to the $\sim$15$\sigma$ detection of PH$_3$ by GRB20 is reproduced, and  is shown to give spurious results. In Section 4, an independent analysis of the ALMA spectrum is presented and discussed. Short conclusions follow in Section 5. 

\section{Processing and Calibration of ALMA data}

The ALMA data processing and calibration is briefly discussed here, but the reader should consult GRB20 for more details. The aim was to perform these in the same way as in the original study, making use of the (updated) scripts provided by GRB20.   This study only concentrates on the high-resolution narrow-band data centred on PH$_3$. 

The raw data of ALMA project 2018.A.00023.S were retrieved from the ALMA Science Archive. The python script Supplementary Software 2 and 3 of GRB20 was used for initial calibration to produce the ALMA data cubes. It selects data from baselines $>$33 m, the range chosen by GRB20 to maximise the  signal-to-noise of the proposed PH$_3$ 1-0 signal, and also forms the main focus of the analysis presented here. The 33 m cut-off is near the second minimum of the visibility amplitudes of the Venus disk at this frequency.  These procedures were  subsequently altered to process the data for different baseline selections, including all baselines, and baselines of $>$20 m and $>$50 m, corresponding to the first and third minimum of the visibility amplitudes. Supplementary Software 4 was used for imaging the data cubes\footnote{Imaging was performed as in GRB20, with mask "circle[[121pix, 96pix],50pix]" and the multiscale clean method (niter 1000000; cycleniter 20000)} Following GRB20, the Venus rest-frame frequency of the PH$_3$ 1-0 transition was adopted to be 266.9445 GHz (M\"uller 2013). The spectral data were binned to velocity steps of 1.10 km sec$^{-1}$.

   \begin{figure}
   \centering
   \includegraphics[width=9cm]{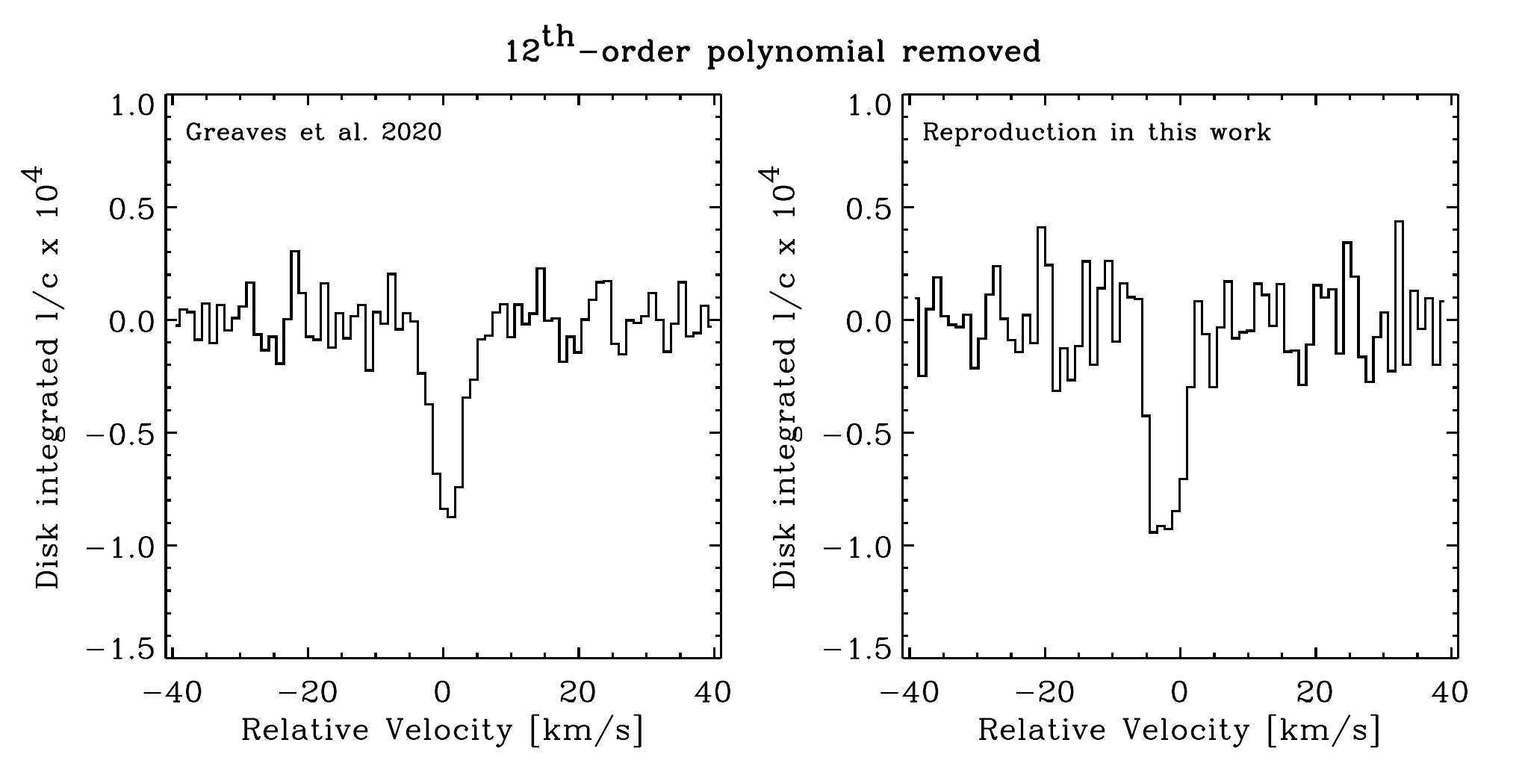}
   \label{reproduction}
      \caption{Reproduction of the ALMA line-spectrum as presented by Greaves et al. 2020, with the original and reproduction in the left and right panel respectively. This is after a 12$^{th}$-order  polynomial is removed from the spectral baseline. The reproduced spectrum is scaled down artificially by a factor 12.8/16.1 to account for the different continuum brightnesses used in the studies. In the reproduction, the line-feature shows a small velocity offset, and the spectral baseline is somewhat more noisy, but the overall signal-to-noise ratio of the two features is similar.}
   \end{figure}

\section{Reproduction of the phosphine results}
At the time of observations, the angular diameter of Venus was 15.36$''$ (GRB20). Since, for the $>$33 m baseline selection, the spectral line data from the limb of the planet still show strong ripples, data from within one major axis of the synthesised beam ($<$1.16$"$) of the planet limb were excluded from analysis. The continuum subtracted line data were summed over the planet disk and divided by the summed continuum data to make the continuum-normalised line-spectrum (l:c). 

To further mitigate the effects of the instrumental ripples and obtain the flattest spectral baseline, GRB20 fitted a 12$^{th}$-order polynomial over a restricted passband of $\pm$40 km s$^{-1}$ around the PH$_3$ transition, interpolating across $|v|<5$ km s$^{-1}$. The central region needs to be masked out, otherwise any line will also be removed.  This procedure was reproduced here. The disk-integrated spectrum obtained by GRB20 is shown in the left panel of Figure 1 with our reproduction in the right panel. Since in this study the continuum level of Venus is found at 12.8 Jy beam$^{-1}$, while it is stated as 16.1 Jy beam$^{-1}$ in GRB20, the reproduced spectrum is artificially scaled down by a factor 12.8/16.1.  The two spectra appear similar, although the line feature is slightly off-centre in the reproduction.  In addition, the reproduced signal is stronger, but the spectrum is also more noisy. The signal-to-noise ratio (SNR)  is estimated by to be $\sim$18 by measuring the peak and standard deviation of the spectrum after applying a boxcar smoothing over 7 velocity steps.  This is very similar (15$\sigma$) to that presented by GRB20. 

   \begin{figure}
   \centering
   \includegraphics[width=9cm]{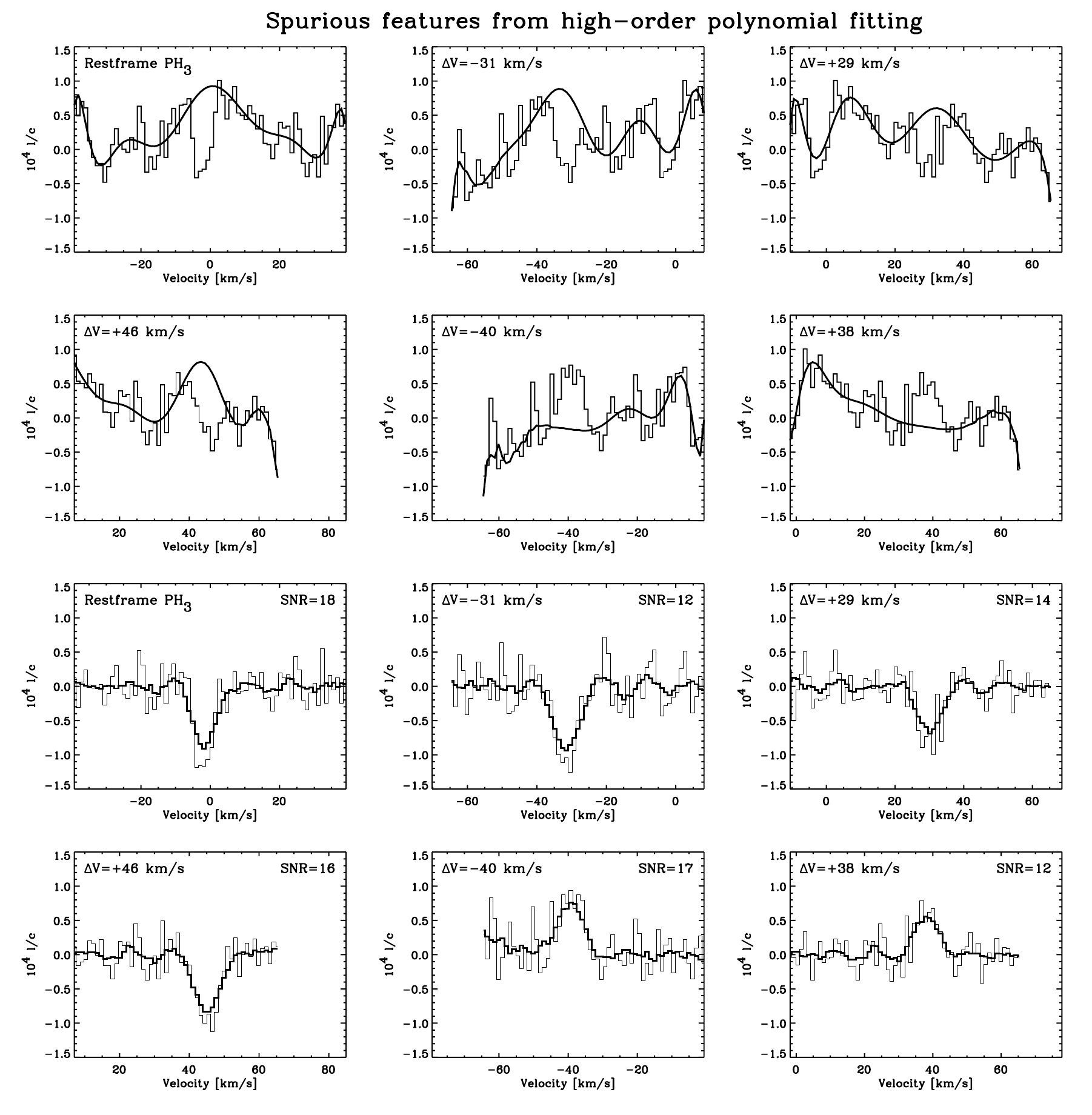}
      \caption{The top two rows show parts of the final ALMA spectrum centred on the transition frequency of PH$_3$ $1-0$ (top left) and five other features, with superimposed the 12$^{th}$-order polynomials fitted to the local data. The bottom two rows show the same with these polynomials removed. 
 We find that now all these features appear at signal-to-noise ratios above 10 within 60 km sec$^{-1}$ of PH$_3$.  It shows that the procedure followed by GRB20 is incorrect, and results in spurious, high signal-to-noise lines.}
         \label{otherfeatures}
   \end{figure}
   
In general, removal of a 12th-order polynomial over a small spectral range in this way has the effect of removing noise structures and instrumental effects. This can lead to severe overestimations of the significance of spectral features and artificial results. To demonstrate this, a search by eye for other features over the observed spectral range of $|v|<60$ km s$^{-1}$ was performed and were subsequently treated with the same procedure. The result is shown in Figure \ref{otherfeatures}. It  leads to at least five other lines with an SNR$>$10, three in absorption and two in emission. The SNR is estimated in the same way as for the feature near the phosphine transition. No plausible 
assignments to the rest frequencies of these features were found. It shows that the procedure followed by GRB20 is incorrect, and results in a spurious, high signal-to-noise line.  

\section{Independent Analysis}
   \begin{figure}
   \centering
   \includegraphics[width=9cm]{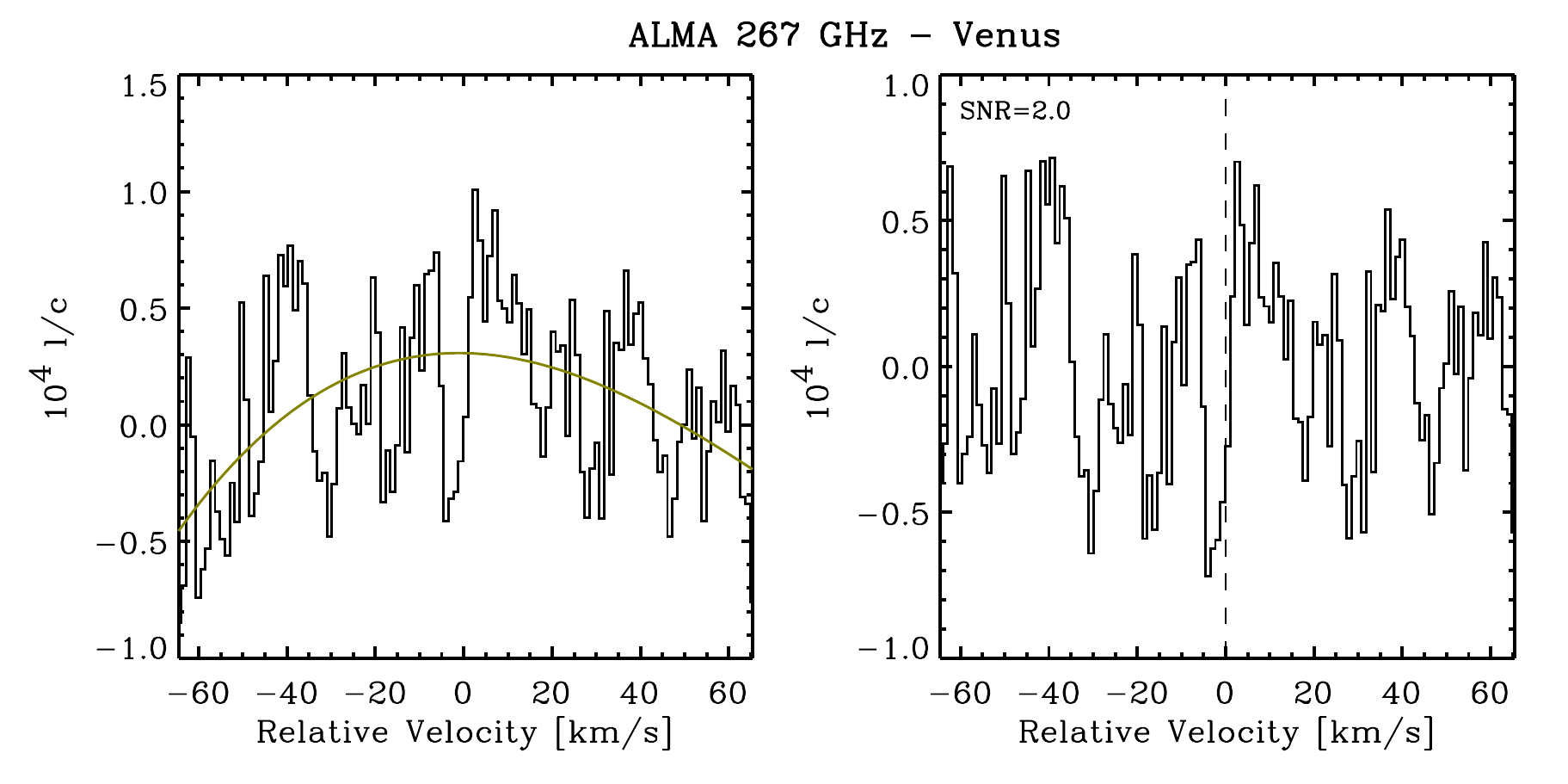}
         \caption{Resulting spectrum from the data analysis presented in this work, before and after the removal of a 3$^{rd}$ order polynomial in the left and right panel respectively. A feature near the  PH$_3$ 1-0 transition is seen at a signal-to-noise of $\sim$2, below the common threshold of statistical significance.}
         \label{independent}
   \end{figure}
   
\begin{figure}
   \centering
   \includegraphics[width=9cm]{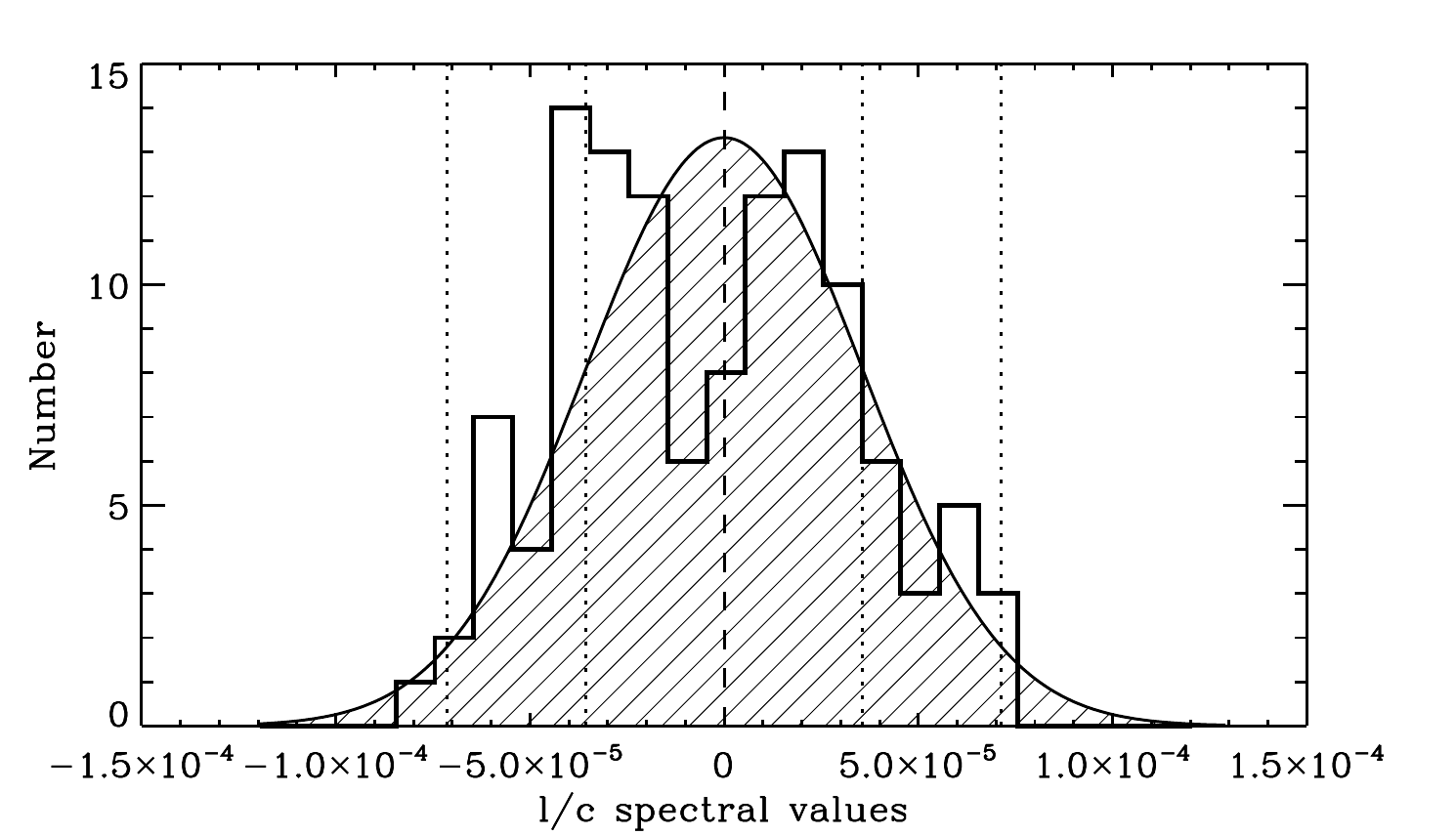}
         \caption{Distribution of the spectral data points of the ALMA 267 GHz observations shown in Figure 3 (histogram) with  the expected Gaussian distribution for the measured standard-deviation of the data overplotted. The vertical dotted lines indicate 1 and 2$\sigma$ limits. The data is clearly non-Gaussian, showing a bimodal distribution,  expected for a spectrum dominated by systematics such as instrumental ripples. This means that low SNR signals can not be reliably linked to a false positive probability.}
         \label{histogram}
   \end{figure}

To independently assess the possible significance of a PH$_3$ 1-0 line in the ALMA data, the disk-averaged l:c spectrum, as shown in the left panel of Figure \ref{independent}, is fitted with a 3$^{rd}$-order polynomial to remove the low-frequency curvature of the spectral baseline. This polynomial is removed from the spectrum as shown in the right panel of Fig. \ref{independent}, resulting in a standard deviation of 3.5$\times 10^{-5}$. The central dip, identified by GRB20 as the PH$_3$ 1-0 line, has an SNR of $\sim$2. Without the polynomial fitting, the SNR is $\sim$1. In astronomy, features at such a low SNR are generally not deemed statistically significant. Furthermore, as is shown in Figure \ref{histogram}, the noise distribution in these data is highly non-Gaussian, as expected for data dominated by systematic ripples. In the absence of other noise factors, systematic effects like sinusoidal and sawtooth ripples can result in extremities at 1.5 $-$ 2 times the standard deviation in the data. This implies that any feature at such levels have no statistical meaning, because they cannot be reliably linked to a false positive probability. 

\begin{figure}
   \centering
   \includegraphics[width=9cm]{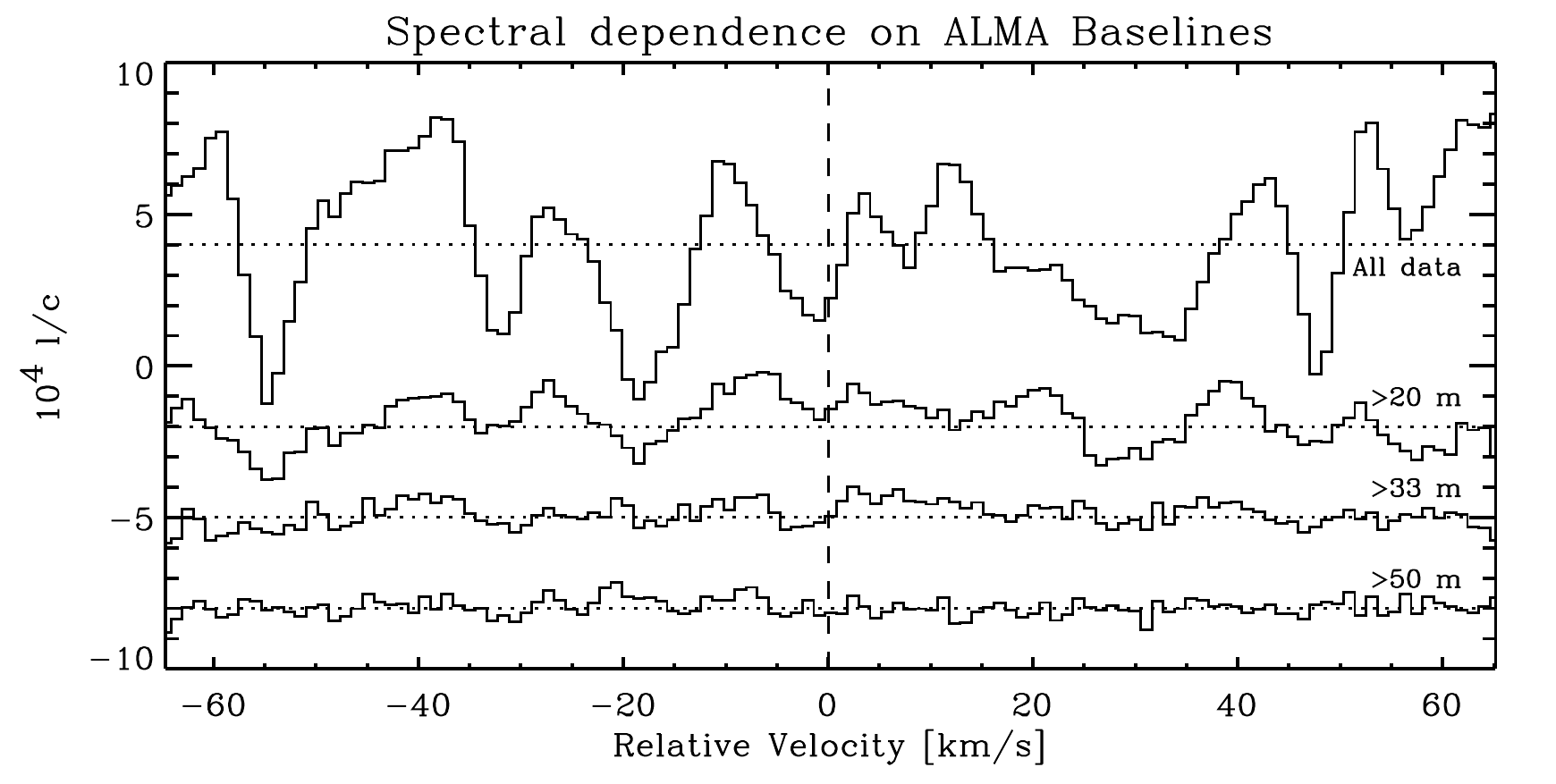}
         \caption{The resulting spectrum for different ALMA baseline selections, vertically offset for clarity, with from top to bottom; all data, $>$20 m, $>$33 m, and $>$50 m baselines. These limits correspond to minima in the visibility amplitudes. Only the spectrum based on the $>$33 m limit exhibits a central dip at an SNR of $\sim$2, implying that this chosen limit has maximised any potential PH$_3$ signal.}
         \label{baselines}
   \end{figure}

As described in Section 2, the ALMA data were calibrated, processed and reduced also using different selections on baseline length. The final disk-integrated l:c spectra are shown in Figure \ref{baselines}, with from top to bottom, for all data and for baselines $>$20 m, $>$33 m (as used for the main analysis), and $>$50 m. These baseline-limits correspond to the first, second, and third minima in the visibility amplitudes of the Venus disk for these observations. In this way, the influence of the adopted baseline limits on the central feature in the spectrum is also assessed.  The spectra based on all data and on a $>$20 m cut-off  show a dip near zero velocity, but at a level that is smaller than several other features in the spectra (SNR$\leq$1). The spectrum based on a $>$50 m cut-off does not show a central feature. Only the spectrum based on the $>$33 m cut-off exhibits a dip at an SNR of $\sim$2, implying that the chosen $>$33 m ALMA baseline-limit has maximised any potential PH$_3$ signal. 

The time dependence of the ALMA spectrum is also investigated by dividing the data in a first and second half. As expected, these spectra based on half of the data are more noisy and provide no evidence for a high-SNR PH$_3$ signal. In addition, the data from Venus' disk was angularly divided into four quadrants, NE, NW, SW, and SE. These spectra are even more noisy and show no candidate features for phosphine. 

\section{Conclusions}
 
 We find that the 267-GHz ALMA observations presented by GRB20 provide no statistical evidence for phosphine in the atmosphere of Venus. The reported $15\sigma$ detection of PH$_3$ $1-0$ is caused by a high-order polynomial fit that suppress the noise features in the surrounding spectrum. The same procedure creates a handful of other $>10\sigma$ lines without plausible spectroscopic assignments, both in emission and absorption, in the direct vicinity of the phosphine 1-0 transition. {\sl Low-order} spectral baseline fitting shows a feature near the expected wavelength at a signal-to-noise of only $\sim2$. While this already in itself is not enough to claim a statistical detection, the noise on the ALMA spectrum is highly non-Gaussian, making any link to a false positive probability unreliable. 
 
 GRB20 provide several arguments to support the validity of their identification of the PH$_3$ feature, including comparison to the JCMT data and a test at offset frequencies. Our analysis, however, shows that at least a handful of spurious features can be obtained with their method, and therefore conclude that the presented analysis does not provide a solid basis to infer the presence of PH$_3$ in the Venus atmosphere.
 
 \begin{acknowledgements}
We thank the authors of GRB20 for publicly sharing their calibration and imaging scripts.
Venus was observed under ALMA Director's Discretionary Time program 2018.A.0023.S. ALMA is a partnership of ESO (representing its member states), NSF (USA) and NINS (Japan), together with NRC (Canada), MOST and ASIAA (Taiwan), and KASI (Republic of Korea), in cooperation with the Republic of Chile. The Joint ALMA Observatory is operated by ESO, AUI/NRAO and NAOJ. 
Data processing was performed at Allegro, the European ALMA Regional Center in the Netherlands. Allegro is funded by NWO, the Netherlands Organisation for Scientific Research.
 \end{acknowledgements}

\end{document}